**Contact transparency in mechanically assembled 2D material devices**

Scott Mills[*†], Naomi Mizuno[*†], Peng Wang[§], Jian Lyu[§], Kenji Watanabe[‡], Takashi Taniguchi[‡], Fernando Camino[ǂ], Liyuan Zhang[§ǂ], Xu Du[†ǂ]

[†]Department of Physics and Astronomy, Stony Brook University, Stony Brook NY 11794-3800, USA

[§]Department of Physics, Southern University of Science and Technology of China, Shenzhen, 518055, China

[‡]National Institute for Materials Science, 1-1 Namiki, Tsukuba 305-0044, Japan

[ǂ] Center for Functional Nanomaterials, Brookhaven National Laboratory, Upton NY 11973, USA

*Abstract*

Two-dimensional atomic crystals (2DACs) can be mechanically assembled with precision for the fabrication of heterostructures, allowing for the combination of material building blocks with great flexibility. In addition, while conventional nanolithography can be detrimental to most of the 2DACs which are not sufficiently inert, mechanical assembly potentially minimizes the nanofabrication processing and preserves the intrinsic physical properties of the 2DACs. In this work we study the interfacial charge transport between various 2DACs and electrical contacts, by fabricating and characterizing 2DAC-superconductor junctions through mechanical transfer. Compared to devices fabricated with conventional nanolithography, mechanically assembled devices show comparable or better interface transparency. Surface roughness at the electrical contacts is identified to be a major limitation to the interface quality.

[*]Equal contribution

[ǂ] Emails: xu.du@stonybrook.edu, zhangly@sustc.edu.cn

The study of two-dimensional (2D) atomic crystals[1] and their heterostructures spans over the past decade from graphene[2, 3] to a variety of semiconductors[4], insulators[5], superconductors[6], topological materials[7, 8], etc. Through the widely practiced approach of mechanical co-lamination/transfer[9, 10], 2D atomic crystals can be directly stacked together with well-adjusted spatial and angular alignment relative to one another[10], even for material combinations which are not possible to synthesize through conventional methods. Such flexible combination of different 2D atomic crystals (2DACs) allows for the discovery of novel devices and artificial materials, which greatly broadens the horizon of low dimensional material research.

An important concern when working with 2DACs and their heterostructures is to minimize the impact of the fabrication process on the intrinsic properties of these materials. Except for a few 2DACs (e.g., graphene, hBN), a majority of the recently explored 2D materials are air-sensitive[10-12]. Hence, the conventional nano-fabrication process often causes degradation to these materials. On the other hand, following the basic approach of co-lamination, device fabrication can also be achieved through mechanical assembly, i.e. direct transfer of 2D materials/heterostructure onto predefined leads. It is therefore important to understand, compared to conventional lithography, how electrically transparent a contact/2DAC interface can be achieved through mechanical assembly, and what the limiting factors to the interface transparency are. In this work, we systematically studied the approach of creating transparent electrical contacts through mechanical assembly, hence minimizing the detrimental effects from conventional nanolithography on some of the air-sensitive 2DACs.

Mechanical assembly of 2DAC devices is based on van der Waals interactions between the 2D crystals and the electric contacts, which may be sufficiently strong so that the interlayer distance is small and charge transfer is facilitated across the interface. For many combinations of

2D crystals, even in presence of surface contamination, the van der Waals force can provide sufficient pressure which pushes contaminants into localized "bubbles", leaving behind atomically clean interfacial areas (the so-called "self-cleaning" mechanism[13]). The possibility of achieving atomically clean interfaces through "self-cleaning" between two 2D layers suggests that it may be possible, in principle, to achieve highly transparent interfaces between 2D layers and electrical contacts. Although this may depend on the surface properties of the contacts, as well as the details of the mechanical assembling process, there appears to be no major fundamental difference between nanolithography/metallization-created and mechanically-assembled metal/2DAC interfaces. Below we discuss metal/2DAC contacts fabricated through mechanical assembly and compare the quality of such contacts with those fabricated with standard nanolithography process. For the characterization of contact quality, superconductor/normal (SN) junctions are studied where interface transparency can be characterized by the bias dependence of the contact resistance. Three types of 2D materials, graphene, $ZrTe_5$ and Bi2212, which cover a wide span of chemical stability, are used for this study.

Figure 1 shows the basic scheme of the mechanical assembly of contacts used in this work. 2D material flakes are exfoliated on top of Methyl methacrylate (MMA)/tape/Polydimethylsiloxane (PDMS) stack prepared on a microscope slide (Duck HD Clear™ Brand tape). The chosen 2D flake is aligned through an optical microscope with the contact electrodes prefabricated on a $SiO_2$/Si substrate. The 2D flake/MMA/tape/PDMS is then brought into contact with the electrodes, followed by substrate heating to ~125-130C. The heating softens/melts the MMA layer so that it gets attached to the substrate once the temperature is lowered, allowing the transfer of both the 2D material and the MMA layer onto the substrate with predefined electrodes. For clarity in discussion, we call the electrical contacts assembled through the above scheme

"press-contacts". Two types of electrodes are used in this work. For normal 2D materials including graphene and ZrTe$_5$, superconducting Nb electrodes are deposited using DC magnetron sputtering on top of 2nm Ti (as adhesion layer), with surface covered by Au (4nm) without breaking vacuum. The Au layer protects the surface of the Nb film from oxidation, but is thin enough to allow for proximity induced superconductivity. For superconducting 2D material Bi2212, normal electrodes Au (25nm)/Ti(2nm) are deposited through electron beam evaporation. We note that for all the samples, the 2D materials are in direct contact with Au. Hence the contact quality is determined by the charge transfer at the Au/2D material interface. The presence of superconducting correlation in our samples provides an effective probe to the interface transparency through the non-ohmic current-voltage characteristics which can be measured by the bias voltage dependence of the differential resistance $dV/dI$[14]. For a highly transparent interface, differential resistance is reduced by the Andreev reflection process and the formation of Josephson supercurrent when the bias voltage is within the superconducting gap. On the other hand, for a low-transparency interface where Andreev reflection and Josephson effect are suppressed, the differential resistance is enhanced due to the suppression of quasiparticle tunneling. Therefore, analyzing the bias voltage dependence of the differential resistance in such SN junctions allows reliable characterization of the interface transparency. In some of the assembly processes (discussed below), to understand the impact of various steps on the quality of the contact, we monitored the two-terminal resistance during the transfer process. The mechanically assembled devices are then measured for their contact resistance and compared to devices fabricated using conventional nanolithography technique.

We first fabricated and characterized mechanically assembled press-contact devices on graphene using superconducting Nb/Au leads. Graphene is arguably the most stable 2D material and shows no degradation in ambient environment. Figure 2A shows the typical differential resistance, dV/dI, as a function of bias voltage measured in such a sample. The normal state resistance (at bias voltage $V < 2\Delta/e$ where $\Delta \sim 1.1$meV is the superconducting gap energy and $e$ is the electron charge) gives the upper limit of the contact resistance, which is ~300Ωμm (per contact). The press-contact resistance at the mechanically-assembled graphene/Au/Nb interface is generally comparable to that in samples made through standard nanofabrication[15, 16].

Compared to superconductor-graphene Josephson weak links fabricated through standard nanolithography method[16], the differential resistance of mechanically assembled structures shows a very different sub-gap ($V < 2\Delta/e$) bias dependence. As illustrated in Figure 2A, the bias-dependent differential resistance here shows a dome-like background which is similar to that in low transparency junctions, where quasiparticle tunneling conductance is suppressed by the superconducting gap[17]. On the other hand, a sharp resistance dip near zero-bias can be observed, which develops deeper with decreasing temperature. This is similar to junctions with a highly transparent SN interface, where Cooper pair tunneling dominates the zero-bias conductance, giving rise to a supercurrent through the Josephson effect[14]. Limited by the relatively high measurement temperature, the Josephson current is smeared by thermal fluctuation (i.e., the thermal excitation $k_B T$ is significant compared to the Josephson energy $E_J = \frac{I_C \phi_0}{2\pi}$, where $I_C$ is the intrinsic Josephson current and $\phi_0$ is the magnetic flux quantum), hence the device resistance does not completely vanish.

A possible scenario which is consistent with such mixed behavior is the presence of a distribution of contact transparencies which contribute in parallel to the total conductance. To qualitatively demonstrate this, in Figure 2B we plot two dV/dI curves taken from two devices made through conventional nanolithography, with high and low transparencies. By adjusting the weight ratio of the two curves, one can calculate the parallel resistance of the devices, with the resulting dV/dI curve showing characteristics qualitatively resembling that in the mechanically assembled devices as shown in Figure 2B (bottom panel). Here we calculate the parallel resistance with a 0.19:1 ratio: $\frac{dI}{dV} = 0.19\frac{dI_1}{dV_1} + \frac{dI_2}{dV_2}$ where 1 and 2 correspond to the high and low transparency junctions, respectively.

Microscopically, the parallel conductance scenario may be understood by considering the surface roughness of the electrodes. The effective area where graphene has high transparency contact with the electrodes is much smaller than the size of the graphene-contact overlap. On the other hand, a significant portion of the contact where graphene "floats" over the electrodes has low transparency. Hence, the overall conductance is the parallel sum of the high and low transparency areas. This scenario is qualitatively supported by the measurement on the surface roughness of the Au/Nb electrodes used in our samples, shown in Figure 2C. Comparing to the roughness of $SiO_2$/Si substrate which is less than 0.5nm, the Au/Nb electrodes show an averaged roughness of over 2nm with high density protrusions which reach over 5nm. The significance of surface roughness in mechanically assembled contacts is further supported by other experiments that will be discussed below.

Next we discuss $ZrTe_5$, which is an air-sensitive layered material. $ZrTe_5$ has been recently studied, in its bulk form, for its chiral magnetic effect as a Dirac semimetal. When exfoliated, $ZrTe_5$ thin films are prone to degradation from reaction with oxygen and humidity. With conventional nanolithography, such degradation results in poor contact quality, as indicated in Figure 3B for a superconductor-normal-superconductor (SNS) junction with Nb (40nm)/Au (4nm) press-contacts. The charge transport across the metal/ZrTe5 interface is largely dominated by hopping[18], which results in a dome-like bias dependent differential resistance with little superconductivity features such as the superconducting gap. For comparison, Figure 3C shows the contact resistance characteristics in a mechanically assembled $ZrTe_5$/Au(4nm)/Nb(40nm)/Ti(2nm) press-contact device (optical image shown in Figure 3A). Significant improvements are observed both in contact resistance and the contact transparency as indicated by the W-shaped bias dependence of the differential resistance which, as discussed below, qualitatively characterizes a SN interface with medium transparency.

To explore the impact of surface roughness on the contact transparency in $ZrTe_5$/Au (4nm)/Nb (40nm) junctions, we compare the contact resistance between mechanically assembled devices and devices fabricated through "via"-contacts[19] (Figure 3A). In the latter case, contact pads are first metalized on perforated hBN flakes on a $SiO_2$/Si substrate. The hBN/contact assembly is then transferred on top of $ZrTe_5$ thin films (see Supplementary Information). Finally, another standard nanolithography is performed to define Au/Ti leads, which connect the "via"-contacts to bonding pads for transport measurements. Since the contact surface is formed on top of $SiO_2$/Si substrate before the transfer, it is much smoother compared to the top surface of thermally evaporated thin films (see Figure 2C). As a result, significantly lower contact resistance is achieved in "via"-contacts compared to that in the press-contacts. On the other hand, the contact transparency, as indicated by the line-shape of the bias-dependent differential resistance, are qualitatively similar in both types of contacts, as shown in Figure 3D. This apparent discrepancy can again be explained considering the surface roughness at the contacts. In press-contacts, the effective contact area is much smaller compared to that in a device with the smooth "via"-contacts, resulting in relatively large contact resistance. On the other hand, at locations where $ZrTe_5$ makes close contact with the metal leads, the local contact transparency is at least comparable to that in the "via"-contacts.

Since we do not observe multiple Andreev reflection nor supercurrent, we can treat the a ZrTe$_5$ SNS device as two SN junctions in series. This allows a more quantitative comparison between the press-contacts and the "via"-contacts using the modified Blonder-Tinkham-Klapwijk (M-BTK)model, which calculates the differential resistance across an SN junction by treating the SN interface as a delta-function potential barrier and considering the life-time broadening effect[20, 21] (see Supplementary Information). Each SN junction is characterized by 3 parameters: the Z factor, which represents the barrier strength at the interface $V(x) = Z\hbar v_F \delta(x)$ ($\hbar$ and $v_F$ being reduced Plank constant and Fermi velocity, respectively); the $\Gamma$ factor, which describes the broadening of the superconducting density of states as a result of inelastic processes as well as a broadened distribution of superconducting gap values; and the effective superconducting gap $\Delta$. Both press-contact and the "via"-contact devices show a barrier strength of Z~1, which characterizes a SN interface with intermediate transparency[17]. However, the "via"-contact device shows a much more significant broadening behavior (large $\Gamma$) and superconducting gap reduction compared to the press-contact device. This may be attributed to the additional lithography step which causes interfacial diffusion between gold and ZrTe$_5$, which both smears and lowers the superconducting gap at the interface. Figure 3A (right panels) presents the cross-section scanning electron microscopy (SEM) images at the contact interfaces for the two devices. Significant and broad diffusion/mixing is visible at the Au/ZrTe$_5$ interface for the "via"-contact. While for the press-contact, the Au layer remains localized and its interfacial diffusion is much weaker. Overall, we found the fabrication of press-contacts to be quick and simple, with contact area limited by the surface roughness of the contact, while the "via" method is more technically involved and is more prone to interfacial diffusion, but creates more intimate contacts with low roughness (hence larger effective contact area) and minimal contact resistance. Both methods create contacts with comparable transparency. In our study here, neither technique could produce a sufficiently transparent interface between Nb/Au and ZrTe$_5$ for multiple Andreev reflection and supercurrent. This may be attributed to the surface degradation of ZrTe$_5$ which was not sufficiently suppressed due to the relatively large residual water vapor and oxygen concentration of our home-built glovebox. We also note that for the nanolithography-made device, the M-BTK model fails to describe the differential resistance behavior in the hopping-dominated charge transport regime.

Finally we fabricated and characterized mechanically assembled devices with high-temperature (HTC) superconductor Bi$_2$Sr$_2$CaCu$_2$O$_{8+x}$ (Bi2212). While generally considered stable in bulk form, Bi2212 may severely react with water at the surface. Conventional nanolithography in the ambient environment fails to create reasonable electrical contact on exfoliated flakes, unless the contact area is plasma/ion-etched through a few unit-cells in-situ before metallization[22]. Stencil/Shadow mask evaporation proves successful in making low-resistance electrical contact with Bi2212[23], which also suggest that conventional lithography process is too harsh for the surface of the material.

Here we assemble Bi2212/Au devices by directly transferring Bi2212 thin flakes onto predefined Au leads, under a relative humidity of ~ 0.1%. Compared to graphene and ZrTe$_5$, we found Bi2212 to be significantly more challenging in forming electrical contact with consistently low resistance. The contact quality appears to be sensitive to the details of the mechanical motion of the flakes under study. To clearly understand the impact of various steps in the assembly procedure, we monitored the two-terminal resistance throughout the process, as shown in Figure 4A.

Upon pressing a Bi2212 flake against the contacts, a low two-terminal resistance interface (typically < 1kΩ) is quickly formed. With a typical contact area of ~few μm$^2$, such contact resistance is comparable to that in devices with standard nanolithography and in situ ion beam contact etching. To transfer the MMA layer together with the Bi2212 flake, the substrate is heated to 130C over a period of ~5 min. With initial heating, the contact resistance slightly decreases, presumably due to softening of MMA and hence better conformation between the Bi2212 flake with the contacts. Further temperature increase, however, causes the contact resistance to increase. Possible reasons for such an increase include: 1) melting of MMA decreases the pressure that presses the Bi2212 against the contacts; 2) surface deterioration from surface chemical reactions with residual water vapor at elevated temperatures. Once the substrate temperature reaches 130C, the heater is turned off and the substrate temperature drops back to below 30C in ~5 min. Over the cooling period the contact resistance decreases slightly but does not completely recover from the increase during the heating period. Finally the PDMS stack is lifted away from the substrate, leaving the MMA layer and the Bi2212 flake transferred to the substrate. A sharp resistance increase has been observed at the moment the PDMS is separated at the location of the samples, "dragging" the Bi2212 flake from the contacts. We found the final step of the assembly to be the most critical in determining the contact quality. With relatively thin MMA (<0.5μm), the "dragging" force from MMA is found to cause a major increase in contact resistance, sometimes even causing the contact to be completely lost. With thicker MMA (>1μm) however, the rigidity of the MMA layer buffers the Bi2212 flake from the dragging force, resulting in rather minor or even no resistance increase at the point of the lift-up. For all successfully assembled devices, we used EL11 MMA spun at 1000 RPM, with a thickness of ~1μm.

In Figure 4B, a mechanically assembled device of Bi2212 thin film ~10nm thick was measured using four-terminal geometry (which measures the thin film resistance in Bi2212 while eliminating the contact resistance), with zero-bias resistance $R$ showing superconducting transition at $T_c$~85K. Figure 4C shows the two-terminal differential resistance of a Au/Bi2212/Au junction. As Bi2212 superconducts below the critical temperature, the measured differential resistance comes solely from the two contacts. The bias-dependence of the differential resistance shows quasiparticle tunneling-like behavior with suppressed density of states at low biases. Unlike in a conventional superconductor junction, the Bi2212/Au press-contacts show a differential resistance which follows ~ 1/V dependence, as a result of the d-wave nature of the superconducting gap. At low temperatures, spikes in differential resistance are observed, which are associated with local break-down of superconductivity. At zero bias we observe a sharp resistance dip. This may be associated with the proximity-induced superconductivity in the contact area, which is suppressed at large bias current.

In summary, we have studied electrical contact transparency between gold and 2D crystals including graphene, $ZrTe_5$ and Bi2212, in a SN junction setup where contact transparency can be characterized through the bias dependence of differential resistance. Several factors are identified to affect the interface quality. For some of the air-sensitive 2D crystals, minimizing the processing in device fabrication (including exposure to heat and ambient) reduces interface degradation and improves the contact transparency. Minimizing heating also reduces interfacial diffusion and improves the sharpness of the metal-2D crystal interface. Through direct transfer of 2D crystals on electrical contacts (press-contact), high contact transparency similar or better compared to those fabricated by conventional nanolithography can be achieved. In such press-contacts, the main factor which limits the contact quality is the surface roughness of the metal leads, which reduces the effective contact area of the device. With further improvement on the surface smoothness of the contacts, it is possible to mechanically assemble 2D crystal-based junctions with high-quality interfaces, expanding such research to a much wider range of novel layered materials.

**Acknowledgement**

X.D. acknowledge support from AFOSR under grant FA9550-14-1-0405. L.Z. acknowledge support from Guangdong Innovative and Entrepreneurial Research Team Program (No.2016ZT06D348), NFSC(11874193) and Shenzhen Fundamental subject research Program (JCYJ20170817110751776) and (JCYJ20170307105434022). K.W. and T.T. acknowledge support from the Elemental Strategy Initiative conducted by the MEXT, Japan and the CREST (JPMJCR15F3), JST. This research used (in part) resources of the Center for Functional Nanomaterials, which is a U.S. DOE Office of Science Facility, at Brookhaven National Laboratory under Contract No. DE-SC0012704.

**Figure Captions**

Figure 1. Schematic illustration of the transfer setup for overlaying 2D crystals with predefined electrical contacts.

Figure 2. Graphene-superconductor interfaces. **A**. Bias voltage dependence of differential resistance for a mechanically assembled graphene-Nb/Au junction. Inset: optical micrograph of the mechanically assembled graphene-Nb/Au junction. The scale bar labels 3μm. **B**. Bias dependence of differential resistance for two graphene-Nb/Au junctions fabricated through conventional nanolithography, with low (top panel) and high (middle panel) interface transparencies. The differential resistance calculated from weighted conductance contribution from the two curves (bottom panel) shows qualitatively similar behavior as that in the mechanically assembled junctions. **C**. Surface roughness comparison between $SiO_2$/Si substrate (left panel) and Au/Nb thin film (right panel), measured using AFM. Both scale bars label 250nm. The histograms show the height distributions.

Figure 3. $ZrTe_5$-superconductor interface. **A**. The left panels show optical micrographs of $ZrTe_5$-Nb/Au junctions with press-contacts (top left) and with "via"-contacts (bottom left). The right panels show corresponding the cross-section SEM images of the two devices, with the cross-section etched with focused ion beam and SEM imaging taken at 52-degree tilt angle. **B**. Bias dependence of the differential resistance for a $ZrTe_5$-Nb/Au junction fabricated through conventional nanolithography. **C**. Bias dependence of the differential resistance for a $ZrTe_5$-Nb/Au junction fabricated by direct transfer of $ZrTe_5$ onto pre-defined Nb/Au leads. The red curve is a fit from the M-BTK model with level broadening. **D**. Bias dependence of the differential resistance for a $ZrTe_5$-Nb/Au junction fabricated through the "via" method. The red curve is a fit from the M-BTK model with level broadening.

Figure 4. Mechanical assembly of Bi2212/Au junction. **A**. Step-by-step monitoring of the Bi2212/Au contact resistance during the assembly process. **B**. Temperature dependence of four-terminal resistance showing superconducting transition. **C**. Bias voltage dependence of the differential resistance from a mechanically assembled Bi2212/Au at various temperatures. From top to bottom, the curves correspond to temperatures: 8K, 15K, 22K, 27K, 39K, 59K and 87K.

Figure 1

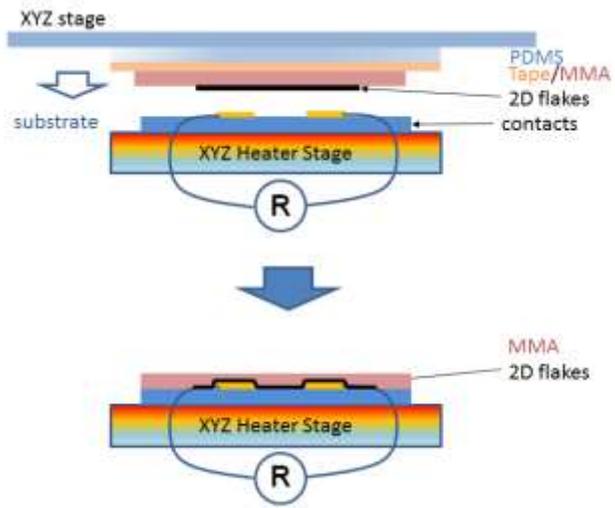

Figure 2

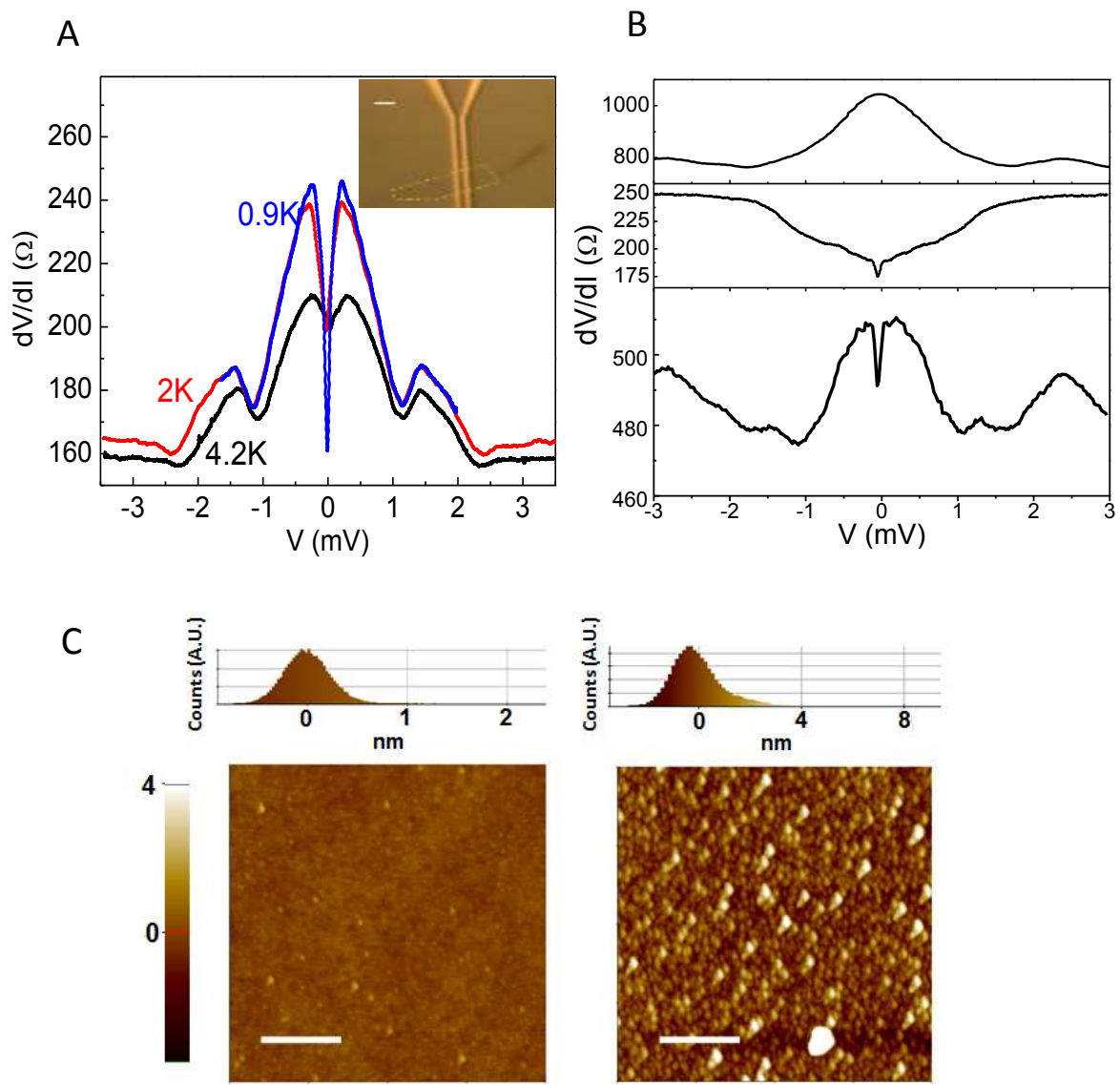

Figure 3

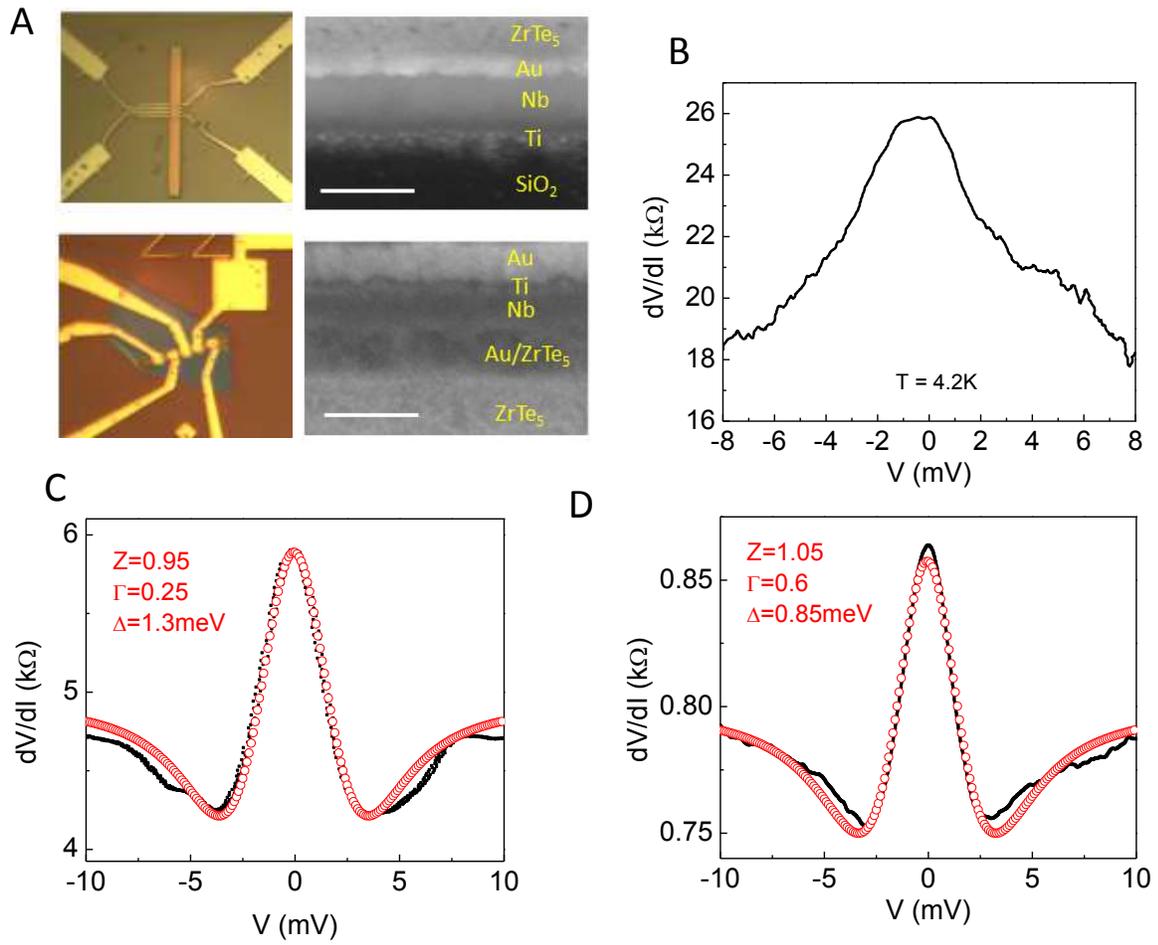

Figure 4.

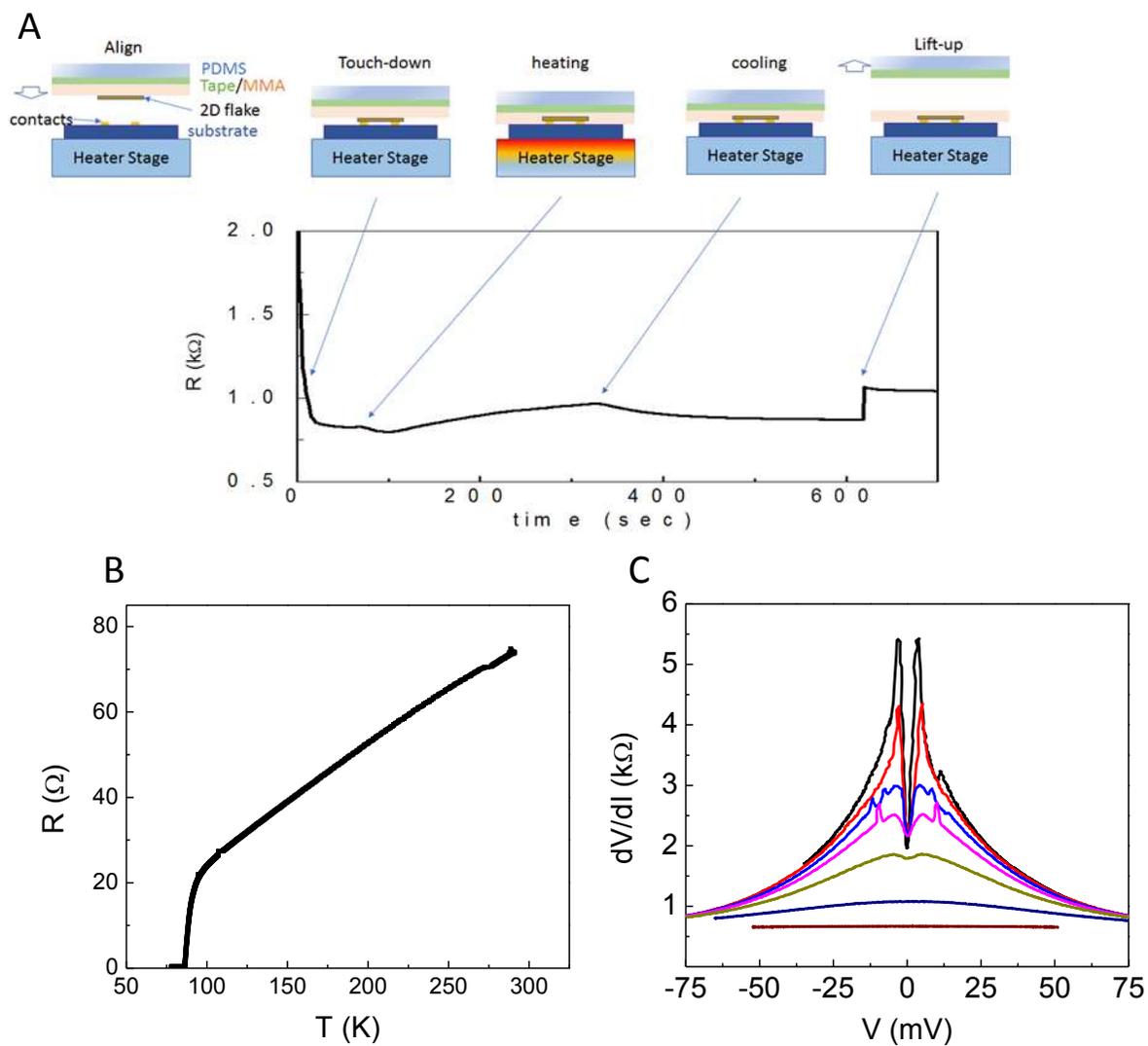